\documentstyle[prd,preprint,aps,epsfig,floats]{revtex}
\begin{document}

\tightenlines
\input epsf.tex
\def\DESepsf(#1 width #2){\epsfxsize=#2 \epsfbox{#1}}
\draft
\thispagestyle{empty}
\preprint{\vbox{\hbox{OSU-HEP-01-05} \hbox{UMDPP-02-003}
\hbox{August 2001}}}

\title{\Large \bf Observable Neutron--Antineutron Oscillations
in Seesaw Models of Neutrino Mass}

\author{\large\bf K.S. Babu$^{(a)}$ and R.N. Mohapatra$^{(b)}$}

\address{(a) Department of Physics, Oklahoma State University\\
Stillwater, OK 74078, USA \\}

\address{(b) Department of Physics, University of Maryland\\
College Park, MD 20742, USA}

\maketitle

\thispagestyle{empty}

\begin{abstract}

We show that in a large class of supersymmetric models with
spontaneously broken $B-L$ symmetry, neutron--antineutron 
oscillations occur at an observable level even though the
scale of $B-L$ breaking is very high, 
$v_{B-L} \sim 2 \times 10^{16}$ GeV, as suggested
by gauge coupling unification and neutrino masses.  
We illustrate this phenomenon in the context of a recently proposed
class of seesaw models that solves the strong CP problem
and the SUSY phase problem using parity symmetry.
We obtain an {\it upper} limit on $N-\bar{N}$ oscillation
time in these models, $\tau_{N-\bar{N}} \leq 10^{9} -
10^{10}$ sec.  This suggests that a modest improvement in
the current limit on $\tau_{N-\bar{N}}$ of
$0.86 \times 10^8$ sec  will either lead to the
discovery of $N-\bar{N}$ oscillations, or will considerably
restrict the allowed parameter space of an interesting
class of neutrino mass models.

\end{abstract}

\newpage

\section{Introduction}

It is widely believed that the most natural and appealing explanation
of the recent neutrino oscillation results is provided by
the seesaw mechanism \cite{seesaw} incorporated into extensions of the Standard Model
that include a local $B-L$ symmetry. The simplest models
with local $B-L$ symmetry are the left-right symmetric models \cite{lr} 
based on the gauge group $SU(3)_C \times
SU(2)_L\times SU(2)_R\times U(1)_{B-L}$.  These models have the 
additional virtue that they explain the origin of parity violation in
weak interactions as a consequence of spontaneous symmetry breaking in
very much the same way as one explains the strength of the weak interaction
in the Standard Model.  Stability of the Higgs sector under radiative
corrections calls for weak scale supersymmetry as in the Minimal Supersymmetric
Standard Model (MSSM).  It has recently been shown that if
the MSSM is embedded into a left--right
symmetric framework at a high scale $v_R \sim 10^{14} - 10^{16}$ GeV,
as suggested by neutrino oscillation data and by gauge coupling
unification, it helps solve  some important problems faced by the MSSM, 
viz., the SUSY CP problem \cite{rasin}, the strong
CP problem \cite{bdmcp} and the $\mu$ problem.  Supersymmetric models with such a
high scale embedding are therefore 
attractive candidates for physics beyond the Standard Model.  

It was noted many years ago \cite{marshak} that the electric charge formula
of the left--right symmetric
models,  $Q= I_{3L} + I_{3R} + \frac{B-L}{2}$, allows
one to conclude from pure group theoretic arguments that parity symmetry
breaking implies a breakdown of $B-L$ symmetry as well
with the constraint that $2\Delta I_{3R} =
-\Delta(B-L)$. This simple relation is profoundly revealing. It says that
the neutrinos must be Majorana particles since the lepton number
breaking terms in the theory must obey $|\Delta L| = 2$ selection
rule.  This conclusion follows directly if Higgs triplets are used to break
$SU(2)_R$ symmetry since $I_{3R} = 1$ for triplets, it also holds 
when Higgs doublets are used for this purpose, since gauge invariance
requires the presence of two such doublets in the mass term for the
neutrinos.  Secondly, for purely hadronic baryon number
violating processes, baryon number must change by at least
two units, $|\Delta B| = 2$. This means that models based
on  left--right symmetric gauge structure  can lead to the
process where a neutron transforms itself into  an antineutron
($N-\bar{N}$ oscillation\cite{kuzmin,glashow,marshak}), while they may
forbid the decay of the proton, which is a $\Delta B = 1$ process.

While the above group theory argument predicts the existence of 
$N-\bar{N}$ oscillation in left--right symmetric models, its strength
will depend on the details of the model.  Using
simple dimensional analysis it is easy to find that the lowest dimensional
operators that contributes to $N-\bar{N}$ oscillation are six quark
operators, a typical one being $(u^cd^cu^cd^cd^cd^c)$.
This operator has dimension 9 and therefore the 
coupling strength  scales as $G_{\Delta
B=2}\sim \frac{1}{M^5}$, where $M$ is the
scale of new physics.  It is natural to
identify $M$ with the scale of $B-L$ (or parity) breaking. 
The current lower limit on $N-\bar{N}$ oscillation time,
$\tau_{N -\bar{N}} \geq 0.86 \times 10^8$ sec \cite{milla},\footnote{This
is the direct
limit from free neutron oscillation searches.  Indirect limits
which involves some reasonable nuclear physics assumptions have
been extracted from nucleon decay experiments which are slightly
more stringent: $\tau_{N-\bar{N}} \geq 1.2 \times 10^8$ sec \cite{kam}.} 
implies an upper
limit $G_{\Delta B=2} \leq 3 \times 10^{-28}$
GeV$^{-5}$.  For $N-\bar{N}$ oscillations to be observable then,
the scale $M$ should be rather low, $M \leq 10^6$ GeV.  

One class of models where $\Delta B=2$
transition manifests itself through Higgs boson exchange
has been discussed in Ref. \cite{marshak}.  
There it was shown that if the $SU(3)_C \times
SU(2)_L\times SU(2)_R\times U(1)_{B-L}$
model is embedded into the $SU(4)_C \times
SU(2)_L\times SU(2)_R$ gauge group,
then $N-\bar{N}$ oscillations can arise at an observable level
if the $SU(4)_C$ breaking scale is in the 100 TeV range.  
In these models, $N-\bar{N}$ oscillation amplitude is intimately tied to 
an understanding of small neutrino masses via the seesaw mechanism as
well as the breaking of quark--lepton degeneracy implied by $SU(4)_C$
symmetry.  The same Higgs field that breaks $SU(4)_C$ and generates
heavy Majorana masses for the right--handed neutrinos also mediate
$N-\bar{N}$ oscillations here.  
With the scale of $SU(4)_C$
breaking in the 100 TeV range, these models would appear to
be incompatible with gauge coupling unification.  
Furthermore, such a low scale
of parity breaking would not yield naturally neutrino masses in the
range suggested by current experiments.  If we raise the scale of
parity/$SU(4)_C$ breaking to values above $10^{12}$ GeV, so that
small neutrino masses in the right range are generated naturally,
then $N-\bar{N}$ transition amplitude becomes unobservably 
small in these models.\footnote{A counter example where a higher
scale of parity violation can go hand in hand with
observable  $N-\bar{N}$ oscillation was
noted in the context of a SUSY $SU(2)_L\times SU(2)_R\times SU(4)_c$ model
in Ref. \cite{chacko}.  These models possess
accidental symmetries that lead to light ($\sim 100$ GeV) diquark
Higgs bosons even though the scale of parity violation is high. As a
result, the $N-\bar{N}$ oscillation operator can have observable strength.
Unification of gauge couplings is however difficult to achieve in
these models.}

Does the above arguments mean that $N-\bar{N}$ oscillations are
beyond experimental reach based on current neutrino oscillation
phenomenology?  
In this Letter we will show that this is not the case in a class
of attractive seesaw models with local $B-L$ symmetry.  
We will see that in these models 
a new class of $\Delta B=1$ operators is induced
as a consequence of parity breaking.  
These operators lead to observable $N-\bar{N}$
oscillation despite the scale $v_R$ of parity breaking being close
to the conventional GUT scale of $2 \times 10^{16}$ GeV.  
In fact, $G_{\Delta B=2}$ increases with $v_R$ and therefore one has the
inverse phenomenon that increasing  $v_R$ leads to stronger
$N-\bar{N}$ oscillation amplitude.  Interestingly, the scale 
$v_R$ implied by neutrino masses is such that $N-\bar{N}$
oscillation should be accessible experimentally with a modest
improvement in the current limit.  We obtain an {\it upper}
limit of $\tau_{N-\bar{N}} \leq 10^8-10^{10}$ sec
in this class of models.  This prediction becomes sharper in a
concrete model where flavor symmetries reduce considerably
the uncertainties in the estimate of $\tau_{N-\bar{N}}$.  
We emphasize that our upper limit is derived in the context
of conventional seesaw models of neutrino mass without using
any special ingredients to enhance $N-\bar{N}$ oscillation
amplitude.  This should provide new impetus for an improved
experimental search for $N-\bar{N}$ oscillations.

\section{High scale seesaw model and $N-\bar{N}$ oscillation}

The basic framework of our model involves the embedding of
the MSSM into a minimal SUSY left--right gauge structure 
at a scale $v_R$ close to the GUT scale.  The electroweak
gauge group of the model, as already mentioned, is
$SU(2)_L\times SU(2)_R\times U(1)_{B-L}$ with the standard assignment of
quarks and leptons --
left--handed quarks and leptons ($Q,L$) transform as doublets of
$SU(2)_L$, while the right--handed conjugate
ones ($Q^c,L^c$) are doublets of $SU(2)_R$.  
The quarks $Q$ transform under
the gauge group as $(2,1,1/3)$ and $Q^c$ as $(1,2,-1/3)$, while the
lepton fields $L$ and $L^c$ transform as $(2,1,-1)$ and $(1,2,+1)$
respectively.  
The Dirac masses of fermions arise through their Yukawa
couplings to two Higgs bidoublet $\Phi_a(2,2,0)$, $a=1-2$.  
The $SU(2)_R\times U(1)_{B-L}$ symmetry is broken down to $U(1)_Y$ in the
supersymmetric limit 
by $B-L=\pm 1$ doublet scalar fields, the right--handed doublet denoted by 
$\chi^c(1,2,-1)$ accompanied by its left--handed partner 
$\chi(2,1,1)$.  Anomaly cancellation requires the presence of
their charge conjugate fields as well, 
denoted as $\bar{\chi^c}(1,2,1)$ and
$\bar{\chi}(2,1,-1)$. The vacuum expectation values (VEVs) 
$\left\langle \chi^c \right\rangle  =
\left\langle \bar{\chi^c} \right \rangle = v_R$ break the left--right
symmetry group down to the  MSSM gauge symmetry. A singlet $S$ is
also used to facilitate symmetry breaking in the SUSY limit.  

It has recently been shown that if there exists a $Z_4$ R symmetry,
the minimal model just described will solve the strong CP problem
and the SUSY phase problem based on parity symmetry \cite{bdmcp}.  Furthermore,
the $\mu$ term will have a natural origin.  One possible $Z_4$
assignment was given in Ref. \cite{bdmcp}.  Here we present a slight
variant, which yields the same superpotential at the renormalizable
level as in Ref. \cite{bdmcp} and thus preserves all its success.  
Under this  $Z_4$, the superpotential $W$ changes sign, as do
$d^2\theta$ and $d^2\bar{\theta}$.  
The quark fields $(Q,Q^c)$ are even, while $(L,L^c)$
transform as $(i,-i)$.  The fields $(\chi, \bar{\chi}, \chi^c,
\bar{\chi^c}, \Phi_a, S)$ are all odd under $Z_4$.

The gauge invariant superpotential consistent with this $Z_4$ R symmetry 
at the renormalizable level is

\begin{eqnarray}
W &=& h_aQ \Phi_aQ^c + h_a'L \Phi_aL^c  + \lambda_a\chi \Phi_a \chi^c +
    \lambda_a'\bar{\chi} \Phi_a \bar{\chi^c} + 
  \nonumber \\
&~& \kappa S (e^{i \xi}\chi^c \bar{\chi^c}+
     e^{-i\xi}\chi \bar{\chi} + aS^2 - M^2)  + \mu_{ab}{\rm
Tr}(\Phi_a\Phi_b) S ~.
\end{eqnarray}
This superpotential breaks the gauge symmetry to that of the Standard
Model in the SUSY limit without leaving any unwanted Goldstone
bosons  and induces realistic quark masses and  mxings.  

The baryon number violating processes as well as neutrino masses arise
in this model from higher dimensional operators induced by Planck scale
physics. They will be the main focus of the rest of the paper.  
We shall pay special attention to the relation between the neutrino
mass and the $N-\bar{N}$ oscillation time. The relevant dimension four
operators in the superpotential
which are scaled by $M^{-1}_{\rm Pl}$ and are
allowed by the $Z_4$ symmetry are:
\begin{eqnarray}
{\cal O}_1 &=& f \left[(L^c\chi^c)^2 +(L\chi)^2 \right]~,\nonumber\\
{\cal O}_2 &=& f'\left[ Q^cQ^cQ^c\bar{\chi^c} + QQQ\bar{\chi} \right]~.
\end{eqnarray}
Operator ${\cal O}_1$ gives rise to Majorana masses for $\nu_R$
of order $v_R^2/M_{\rm Pl}$.  Combining this with the Dirac neutrino
masses arising from Eq. (1), small neutrino masses will be generated
by the seesaw mechanism.  
For $v_R \sim 10^{14}-10^{16}$ GeV, the
magnitude of the light neutrino masses are in the right range to explain
the atmospheric and the solar neutrino oscillation data.  Operator
${\cal O}_2$, which is also invariant under the $Z_4$,
leads to baryon number violation.  While
${\cal O}_{1,2}$ could
have their origin in quantum gravity, they  may also be induced by integrating
out vector states that have $Z_4$--inviariant
masses of order the Planck scale.

Note that operators such as $L \Phi \chi^c, L L L^c \chi^c, QLQ^c\chi^c$
are not allowed by the $Z_4$ symmetry.  If they were present along
with ${\cal O}_2$, they would lead to rapid proton decay.  
Note also that the well known proton decay operator $QQQL$ is not allowed by the
$Z_4$ symmetry. In any case its presence would not have been a problem
since it is scaled by the Planck mass and therefore can lead to a proton
lifetime consistent with the present lower limit.

To see the connection between neutrino masses and the
$N-\bar{N}$ oscillation time $\tau_{N-\bar{N}}\equiv (1/\delta
m_{N-\bar{N}})$ qualitatively, first we note that the operator ${\cal O}_1$
leads to the Majorana mass for the right handed neutrino $M_R=
\frac{fv^2_R}{M_{\rm Pl}}$. 
The seesaw formula then leads to the
relation $m_{\nu}=
\frac{M_{\rm Pl}m^2_{\nu^D}}{fv^2_R}$. On the other hand, the operator
${\cal O}_2$ leads to a $\Delta B=1$ operator with strength
$\frac{v_R}{M_{\rm Pl}}$. Leaving aside the details of the flavor structure
of ${\cal O}_2$ and how actually $\delta m_{N-\bar{N}}$ arises, it is
clear that we have a simple linear relation between the neutrino masses
and the $N-\bar{N}$ oscillation time:
\begin{eqnarray}
m_{\nu} = C\frac{\tau_{N-\bar{N}}}{M_{\rm Pl}}
\end{eqnarray}
where  $C$ is a dimensional constant which depends only on the details of
weak scale physics and does not involve the high scale $v_R$. We will 
evaluate $C$ in the next section. This simple relation makes it clear that
our present knowledge of the neutrino masses allows a direct prediction of
the $N-\bar{N}$ oscillation time in the context of the supersymmetric
left-right models broken by doublet Higgs fields.

\section{From supersymmetric $\Delta B=1$ operator to $N-\bar{N}$
oscillations}

Let us now proceed to examine the expected $N-\bar{N}$ oscillation
time resulting from the $\Delta B=1$
operator ${\cal O}_2$ of Eq. (2).
An important point to note here
is that since ${\cal O}_2$ is a superpotential term with antisymmetric
color contraction it must have antisymmetric flavor contraction as well.  The
flavor structure of this operator is then of the
type $u^cd^cs^c$, $u^cd^cb^c$ or $u^c s^c b^c$ in terms of the superfields. 
We must then use flavor mixings 
to obtain the fermionic operator of the type $u^cd^cd^c$ and
then the six quark $N-\bar{N}$ operator. 
The dominant contribution to this process comes from the Feynman diagram
shown in Fig. 1 which proceeds through the exchange
of a gluino and squarks \cite{zwirner}
and involves two $\tilde{d^c}-\tilde{b^c}$ mixings. 
The strength of the $\Delta B=2$ operator resulting from Fig. 1 can be
estimated to be
\begin{eqnarray}
G_{\Delta B=2}\simeq
{2 g_3^2 [(\delta^{13}_{RR})]^2f'^2 \over {M_{\tilde{g}} m_{\tilde{q}}^4}}~,
\end{eqnarray}   
where $M_{\tilde{g}}$ is the gluino mass, $m_{\tilde{q}}$ is the squark
mass and $(\delta^{13}_{RR})$ is the $\tilde{d^c}-\tilde{b^c}$ mixing
angle.  The effective baryon number violating $u^cd^cb^c$ 
Yukawa coupling in the superpotential
is parametrized here as $f'(v_R/M_{\rm Pl})$ (see Eq. (2)).

\begin{figure}[htb]
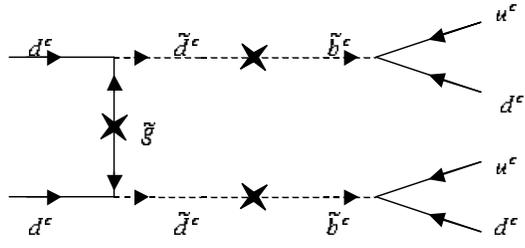

\centerline{ \DESepsf(fig1nnbar.epsf width 8 cm) }
\bigskip
\bigskip
\caption {\label{fig1} Tree level gluino--squark diagram for $N-\bar{N}$
oscillations.}
\end{figure}

Let us first discuss the origin of the flavor mixing that changes a
$u^cd^cb^c$ operator to the required $u^cd^cd^c$ operator.
The dominant source for this in the present context turns out
to be  the mixing of
$\tilde{b^c}$ with $\tilde{d^c}$.  Such mixings occur in the left--right
supersymmetric model since the right--handed quark mixings are
physical above the scale $v_R$.  The renormalization group evolution
of the soft SUSY breaking mass parameters between $M_{\rm Pl}$
and $v_R$  will then induce 
mixings in the right--handed down squark sector proportional to the
top--quark Yukawa coupling and the right--handed CKM mixings.  This is
analogous to the RGE evolution in
the MSSM inducing squark mixing in the left--handed down squark sector
proportional to the left--handed CKM angles 
and the top--quark Yukawa coupling.  We estimate this
right--handed $\tilde{d^c}-\tilde{b^c}$ mixing to be
\begin{equation}
(\delta^{13}_{RR}) \simeq {
\lambda_t^2 (3m_0^2+A_0^2) \over 8\pi^2(m_0^2+8M_{1/2}^2)}
(V_{td}^*V_{tb})~{\rm ln}(M_{\rm Pl}/v_R)
\simeq 2 \times 10^{-4}~.
\end{equation}
This estimate is obtained by integrating out the RGE between $M_{\rm Pl}$
and $v_R$ assuming universality of masses at $M_{\rm Pl}$ \cite{martin}.  
Since above 
$v_R$, both $t^c$ and $b^c$ are part of the same
$SU(2)_R$ multiplet, unlike in the MSSM, $b^c$ Yukawa coupling
is of order one.
In this momentum range, the top--quark Yukawa coupling reduces the mass
of $\tilde{b^c}$.  
In going to the
physical basis of the quarks, this effect will induce the squark mixing quoted in 
Eq. (5).  For the numerical
estimate we took $m_0 = M_{1/2}$ and $A_0=0$ and $v_R \sim
2 \times 10^{16}$ GeV for illustration.  

There is a second source of flavor violation that induces $\tilde{d^c}-
\tilde{b^c}$ mixing in general SUSY models.  That is the baryon number
violating Yukawa couplings themselves.  If we write in  standard
notation, the effective $B$--violating 
superpotential arising from Eq. (2) as $W \supset (\lambda''_{ijk}/2) 
u^c_i d^c_j d_k^c$, the RGE evolution from Planck scale
to the weak scale will induce $\tilde{d^c}-\tilde{b^c}$ mixing 
proportional to $\lambda''_{ijk}$.  For example, if we keep only  the couplings
involving the $u^c$ quark, viz., $\lambda''_{123},
\lambda''_{113}$ and $\lambda''_{112}$, we can estimate the induced
$(\delta^{13}_{RR})$ by integrating  the relevant RGE \cite{white} to be
\begin{equation}
(\delta^{13}_{RR}) \simeq {\lambda''_{121} \lambda''_{123} \over 4 \pi^2}
 {(3m_0^2 + A_0^2) \over (m_0^2+8M_{1/2}^2)}{\rm ln}(M_{\rm Pl}/M_Z)~.
\end{equation}
Recalling that $\lambda'' \sim v_R/M_{\rm Pl}$, we see that while
this source of flavor mixing may not be negligible, it would be 
typically smaller than the ones from the right--handed quark mixings
of Eq. (5).

A third source of flavor violation relevant for $N-\bar{N}$
oscillations has been identified in Ref. \cite{sher} involving
the exchange of the Wino.  Such diagrams will have an electroweak
loop suppression and a chirality suppression necessary to convert
the left--handed squark to the right--handed one.  We find that
this contribution to $\delta m_{N-\bar{N}}$
has a suppression factor given approximately
by $[(\alpha_2/4 \pi) (m_b/m_{\tilde{q}})]^2
\sim 1 \times 10^{-9}$ (valid for small $\tan\beta$)
which is about two orders of magnitude smaller in this class of models
compared to the gluino exchange diagram of Fig. 1.  

One has to calculate the hadronic matrix element of the six quark operator
in order to obtain the $\tau_{N-\bar{N}}$. This has been discussed in
several places in the literature\cite{rao}. The calculations of this
``conversion'' factor can be done using crude physical arguments,
according to which one has to multiply the $G_{\Delta B=2}$ by
$|\psi(0)|^4$ to obtain $\delta m_{N-\bar{N}}$  where $\psi$ is the
baryonic wave function for three quarks inside a nucleon. On dimensional
grounds, one can deduce that $|\psi(0)|^4\simeq \Lambda^6_{QCD}$, which
implies that $\delta m_{N-\bar{N}}\sim 10^{-5}G_{\Delta B=2}$ GeV.
More detailed bag model calculations have been carried out.  Rao and
Shrock in Ref. \cite{rao} quote this conversion factor to be
$2.5 \times 10^{-5} G_{\Delta B = 2}$.  We shall use this number for our numerical
illustrations.  

Combining this matrix element with Eq. (4)-(5) we obtain
\begin{equation}
\tau_{N-\bar{N}} \simeq {7 \times 10^8 sec.  \over f'^2} \left({2 \times 
10^{14} ~{\rm GeV}
\over v_R}\right)^2 \left({M_{\tilde{g}} \over 500 ~{\rm GeV}}\right)
\left({m_{\tilde{q}} \over {500 ~{\rm GeV}}}\right)^4~.
\end{equation}

We can rewrite Eq. (7) in a form that makes the connection with the
neutrino mass more transparent.  The mass of $\nu_\tau$ can be expressed
through the seesaw formula from Eq. (2) as $m_{\nu_\tau} = (m_{\nu_\tau^D})^2
M_{\rm Pl}/(f v_R^2)$ where $m_{\nu_\tau^D}$ denotes the
Dirac mass of $\nu_\tau$.  Eliminating the high scale $v_R$ from this, we
have from Eq. (7),
\begin{equation}
\tau_{N-\bar{N}} \simeq 2.8 \times 10^4 sec. \left({f \over f'^2}\right)
\left({m_{\nu_\tau} \over
0.06 ~{\rm eV}}\right) 
\left({m_t \over m_{\nu_\tau^D}}\right)^2
\left({M_{\tilde{g}} \over 500~ {\rm GeV}}\right)
\left({m_{\tilde{q}} \over 500~ {\rm GeV}}\right)^4~.
\end{equation}

Since the value of $m_{\nu_{\tau}}$ can be determined from the atmospheric
neutrino data under certain assumptions, we conclude that 
within the seesaw 
framework, measurement of $N-\bar{N}$ oscillation will be a measure of the
Dirac mass of the tau neutrino. This can then be used as a way to
discriminate between models of neutrino masses.

To see the specific prediction for $N-\bar{N}$ oscillations
within the context of the class of models
under consideration, we need to know the $\nu_\tau$ Dirac mass.  We can
estimate it from the 
following relations for the Dirac masses of the third generation quarks 
and leptons in the SUSY left--right model:

\begin{eqnarray}
m_t &=& (h_{t,1} \cos\alpha_u + h_{t,2} \sin\alpha_u) v_u \nonumber\\
m_b &=& (h_{t,1} \cos\alpha_d + h_{t,2} \sin\alpha_d) v_d \nonumber \\
m_{\nu_\tau^D} &=& (h_{\tau,1} \cos\alpha_u + h_{\tau,2} \sin\alpha_u)v_u
\nonumber \\
m_{\tau} &=& (h_{\tau,1} \cos\alpha_d + h_{\tau,2} \sin\alpha_d)v_d
\end{eqnarray}
Here $\alpha_{u,d}$ are the Higgs mixing parameters obtained from
Eq. (1) (Eg: $\tan\alpha_u = \lambda_1/\lambda_2$) and $v_{u,d}$ are
the VEVs of the MSSM doublets.  From Eq. (9) it follows that in the
limit of $h_{t,1} \gg h_{t,2}$ and $h_{\tau,1} \gg h_{\tau,2}$,
we get $m_{\nu_\tau^D} \simeq m_\tau {m_t \over m_b}$.  Since at
such high scales $m_b \simeq m_\tau$, this predicts $m_{\nu_\tau^D} \simeq
m_t$.  In fact, we find that unless the two terms in Eq. (9) for
$m_{\nu_\tau^D}$ are precisely canceled, the Dirac mass of $\nu_\tau$
will be approximately equal to $m_t$.  

Using $m_{\nu_\tau^D} \simeq m_t$ and $f\sim 1$, $f' \sim 10^{-1}$, we get a value
for the $N-\bar{N}$ oscillation time which is tantalizingly close to
the present experimental lower limit \cite{milla}.  
For values of $m_{\nu^D_{\tau}}$ 10 times
smaller than $m_t$, and taking the supersymmetric particle masses as
large as 1 TeV, we see that
$\tau_{N-\bar{N}}$ is less than $9 \times 10^9$ seconds,\footnote{We have
not included the QCD evolution factor from the SUSY scale of few hundred
GeV to the GeV scale. Based on the QCD factor of 1.33 for the three quark
proton decay operator \cite{nihei}, we estimate that the corresponding factor
for the $N-\bar{N}$ case should be about 2, which would reduce the
estimate of $\tau_{N-\bar{N}}$ by a factor of 2.}
which is in the range accessible to a recently proposed experiment
\cite{kamys}. It would thus appear that a search for neutron-antineutron
oscillation will provide an enormously useful window into neutrino mass
models and as such a powerful constraint on the nature of new physics
beyond the Standard Model. 

The prediction for $N-\bar{N}$ oscillations can be sharpened if we
make use of flavor symmetries to determine the coefficients $f$ and
$f'$ in Eq. (8).  We illustrate this with a specific choice of flavor
symmetry \cite{bm} taken to be $SU(2)_H \times U(1)_H$.  The first
two families of fermions form doublets of $SU(2)_H$ and have a
$U(1)_H$ charge of $+1$ while the third family fermions are singlets
under both groups.  This flavor symmetry is broken by
a pair of doublets $\phi(1) + \bar{\phi}(-1)$ and singlets $\chi(1)+
\bar{\chi}(-1)$.  Allowing for effective operators suppressed by
a scale $M$ larger than the VEVs of these fields provides a
natural explanation of the fermion mass and mixing angle hierarchy.  If we choose
$\left\langle \phi \right\rangle = \left\langle \bar{\phi} \right\rangle
 = \epsilon_\phi M$ and $\left\langle \chi \right\rangle = \left\langle
\bar{\chi}\right\rangle = \epsilon_\chi M$, a resonable fit to all
quark and lepton masses is obtained, including neutrinos, for
$\epsilon_\phi \simeq 1/7$ and $\epsilon_\chi \simeq 1/20$ and all
dimensionless couplings being order one \cite{bm}.  In this model, we can estimate
the couplings $f$ and $f'$ from the horizontal quantum numbers.
They are  $f \sim \epsilon_\phi^2$ and
$f' \simeq \epsilon_\chi^2 \sin \theta_C$, so 
that\footnote{Normally, the parameter $f$ could have been of
order one but in the horizontal model or Ref. \cite{bm}, due to large
$\nu_{\mu}-\nu_{\tau}$ mixing, it is the $\nu_{\mu}$ flavor entry
that dominates the atmospheric neutrino mass difference and hence
the horizontal suppression factor $\epsilon_{\phi}^2$. The
$\epsilon^2_{\chi}$ factor is due the fact that the operator must
be invariant under $U(1)_H$.} $f/f'^2 \simeq
8 \times 10^4$  This
estimate
leads to $\tau_{N-\bar{N}} \simeq 2 \times 10^9$ sec from Eq. (8).  
Allowing for uncertainties of order 1 in this estimate, we 
expect that $\tau_{N-\bar{N}}$ not
to exceed about $10^{10}$ sec.

Before we conclude a few comments are in order:

\begin{enumerate}

\item The model becomes unacceptable as soon as $SU(3)_C\times U(1)_{B-L}$
is embedded into a higher symmetry such as
$SU(4)_C$ or SO(10) group because in that case, the
$\Delta B=1$ operator described in Eq. (2) is accompanied by other R-parity
violating operators coming from the same higher dimensional operator
${\cal O}_{1,2}$ due to the higher symmetry. Together, they would lead to
unacceptable proton decay rate. Thus observation of $N-\bar{N}$
oscillation would be a signal of an explicit $SU(3)_C\times U(1)_{B-L}$
symmetry all the way upto  the Planck (or string) scale.

\item Baryogenesis has to proceed through a weak scale scenario since
the $\Delta B=1$ interactions in the model are in equilibrium down to the
TeV scale and will wash out any primordial baryon or lepton asymmetry.
We note that the baryon number violating interactions contained in
${\cal O}_2$ themselves can
potentially be the source of weak scale baryogenesis \cite{sarkar}.

\item The lightest neutralino in this model is unstable and will decay via
$\chi^0\rightarrow qqq$ modes due to the presence of the effective $\Delta
B=1$ operator ${\cal O}_2$.  This prediction is directly testable at
colliders.  An alternative candidate for dark
matter must be sought.

\end{enumerate}

\section{Conclusions}

In conclusion, we have found that in a large class of seesaw models for neutrino
masses, despite the high scale of sessaw dictated by the current neutrino
oscillation data, neutron--antineutron oscillation is in the observable range. In
fact, unless the Dirac masses of neutrinos are far below those deduced
under simple and reasonable assumptions, we predict an upper bound on the
neutron-antineutron oscillation time in the range of $10^{9}-10^{10}$ sec. 
This is very close to the present experimental lower limit on $N-\bar{N}$
oscillations.  
In the most conservative theoretical scenario, the measurement of $N-\bar{N}$
oscillation time would be a measure of the Dirac mass for the tau
neutrino, given the values of squark masses.  This in itself would be an
extremely interesting result, since it would discriminate among theoretical
models of neutrino masses.  This is apart from the fundamental importance
that any observation of baryon number violation will carry.  
We therefore strongly urge a new experimental
search for neutron--antineutron oscillation. 

\section*{Acknowledgements}

The work of KSB has been  supported in part by DOE Grant 
\# DE-FG03-98ER-41076, a grant from the Research Corporation,
DOE Grant \# DE-FG02-01ER4864 and by the OSU Environmental 
Institute. RNM is supported by the National Science Foundation 
Grant No. PHY-0099544.  KSB is grateful to the Elementary
Particle Theory Group at the University of Maryland for
the warm hospitality extended to him during a visit when 
this work was completed. One of the authors (R. N. M.) would like to thank
Y. Kamyshkov and R. Shrock for discussions.

\end{document}